\documentclass[Afour,sageh,times]{sagej}

\usepackage{moreverb,url}
\usepackage[colorlinks,bookmarksopen,bookmarksnumbered,citecolor=red,urlcolor=red]{hyperref}

\newcommand\BibTeX{{\rmfamily B\kern-.05em \textsc{i\kern-.025em b}\kern-.08em
T\kern-.1667em\lower.7ex\hbox{E}\kern-.125emX}}

\def\volumeyear{2016}

\begin{document}

\runninghead{Xia et al.}

\title{Comparative Analysis Vision of Worldwide AI Courses}

\author{Jianing Xia\affilnum{1}, Man Li\affilnum{2} and Jianxin Li\affilnum{1}}

\affiliation{\affilnum{1}Deakin University, Australia\\
\affilnum{2}Macquarie University, Australia}

\corrauth{Jianing Xia, Man Li, Jianxin Li}

\email{jianxin.li@deakin.edu.au}

\begin{abstract}
This research investigates the curriculum structures of undergraduate Artificial Intelligence (AI) education across universities worldwide. By examining the curricula of leading universities, the research seeks to contribute to a deeper understanding of AI education on a global scale, facilitating the alignment of educational practices with the evolving needs of the AI landscape. This research delves into the diverse course structures of leading universities, exploring contemporary trends and priorities to reveal the nuanced approaches in AI education. It also investigates the core AI topics and learning contents frequently taught, comparing them with the CS2023 curriculum guidance to identify convergence and divergence. Additionally, it examines how universities across different countries approach AI education, analysing educational objectives, priorities, potential careers, and methodologies to understand the global landscape and implications of AI pedagogy.
\end{abstract}

\keywords{artificial intelligence education,  curriculum structures, educational data mining, AI courses, content analysis, vision analysis}

\maketitle

\section{Introduction}
In recent years, artificial intelligence (AI) has pervaded our daily lives, driven by high-impact machine learning (ML) applications. 
It plays integral roles in finance, healthcare, transportation, and law enforcement, alongside various applications in computer science. Additionally, AI techniques are increasingly utilized in fields such as biology, chemistry, physics, art, and architecture, addressing complex challenges across diverse domains \cite{10371554, DoughtyWBQWZZDS24, williamson2024, Nithithanatchinnapat24, Xu24}.
Concepts like ``neural networks" and ``deep learning" have gained widespread recognition, enabling individuals to perform tasks previously reserved for experts.

In an era marked by rapid advancements in artificial intelligence (AI), the pivotal role of education in preparing the next generation of AI professionals has become increasingly evident \cite{Zhou2021, Brusilovsky24, ZhouZLWLZ22}. 
Given the prevalence of AI, it has naturally become a central focus in the education of computer scientists, and as AI continues to permeate various aspects of society, the vision and strategies adopted by universities in shaping AI education programs hold profound implications for the future workforce and innovation landscape \cite{Wang23, Druga22, Gibellini23}. 
However, the fast pace of AI developments poses a significant challenge in crafting a curriculum that keeps up with the latest advancements over an extended period.

Recognizing the growing importance of Artificial Intelligence (AI) in modern technology, the Association for Computing Machinery (ACM) and the Institute of Electrical and Electronics Engineers – Computer Society (IEEE-CS) collaborated with the Association for the Advancement of Artificial Intelligence (AAAI) to integrate AI more comprehensively into the latest revision of their curricular guidelines, known as CS2023 \cite{EatonE24}. Within the AI Knowledge Area (KA), twelve knowledge units were identified, with five units partitioned among the Computer Science (CS) Core, the KA Core, and electives (Fundamental Issues, Search, Fundamental Knowledge Representation and Reasoning (KRR), Machine Learning, Applications, and Societal Impact).
Given these significant updates to AI education standards, understanding and comparing the curriculum structures of AI education across universities worldwide becomes imperative. My research aims to analyze these curricula to identify best practices, gaps, and opportunities for improvement, ensuring that AI education remains aligned with industry demands and technological advancements.

By systematically gathering information on undergraduate AI programs and related courses from a select group of leading universities, this study aims to provide a comprehensive overview of AI education at the undergraduate level, shedding light on the curriculum structures, course offerings, and educational priorities shaping the next generation of AI professionals.
Drawing from our research objectives, our research aims to address several following questions:

\begin{enumerate}
\item Analysis of Curriculum Structures:
What about the diverse course structures adopted by leading universities in the field of artificial intelligence education? We delve into the contemporary prospects of AI through a comprehensive analysis of the diverse course structures adopted by leading universities. Through this examination, we aim to understand the prevailing trends, methodologies, and priorities in AI education, revealing the nuanced approaches taken by different institutions.
\item Identification of Core AI Topics:
Which AI-related topics and learning contents are most frequently taught? Through rigorous analysis, we seek to identify the foundational pillars of AI education and compare them with the AI knowledge areas outlined in the CS2023 curriculum guidance.
By comparing prevalent educational practices with the prescribed AI knowledge areas guidance, we aim to discern areas of convergence and divergence, facilitating a nuanced understanding of AI pedagogy.
\item Comparative Analysis of AI Education:
How do universities from different countries approach AI education? We embark on a comparative exploration,
analyzing the educational objectives, priorities, potential careers and methodologies across diverse geographical locations. By examining similarities and differences, we
aim to elucidate the global landscape of AI education
and its implications.
\end{enumerate}
By addressing these research questions, our research strives to contribute to a deeper understanding of AI education, its evolving contours, and its global impact.
\section{Methodology}
This section includes both data collection and analytical metrics used in the study, providing a comprehensive approach to understanding the educational landscape of AI programs across selected universities.
\subsection{Data Collection}

Utilizing insights from both the ``2024 AI Index Report" by Stanford University \cite{stanford2024} and the ``QS World University Rankings by Subject 2024: Data Science and Artificial Intelligence \cite{qs2024}," a rigorous selection process was undertaken to identify leading universities at AI education. Institutions from nine countries spanning four continents are included for detailed examination. These regions encompass the United States and Canada in North America, the United Kingdom and Italy in Europe, China (encompassing both mainland China and the Hong Kong Special Administrative Region), Singapore, Malaysia, and India in Asia, in addition to Australia in Oceania.

The data collection methodology for this study involved a thorough investigation of university websites, with a specific focus on institutions listed in the QS World University Rankings \cite{qs2024}. Among the 72 universities distinguished for their prowess in these domains, particular emphasis was placed on identifying universities offering undergraduate programs in Artificial Intelligence (AI). 

The selection criteria prioritized universities that explicitly offered bachelor's degree programs specializing in AI or closely related fields. Additionally, universities with prominent AI research departments or interdisciplinary programs incorporating AI education for undergraduates were included in the data collection process.

Each university's official website served as the primary source of information for gathering data on the curriculum structure and course offerings. For universities offering undergraduate AI programs, the units comprising the curriculum were meticulously documented. In cases where AI was not explicitly offered as a standalone bachelor's degree program, relevant courses closely aligned with AI principles, such as machine learning, data mining, or computer science courses with significant AI content, were identified and included in the dataset.

The data collection process involved thorough navigation of university websites, exploration of program descriptions, examination of course catalogs, and consultation of academic departments or faculty profiles where necessary. Special attention was paid to ensuring the accuracy and completeness of the collected data, focusing on capturing the diverse settings and approaches to AI education across different institutions.

\subsection{Analytical Metrics}
In the following section, we employ various analytical techniques to gain insights into our dataset of course names. We utilize methods such as word clouds to visualize popular words in course names, bar charts to depict the distribution of courses across different categories and the popularity of various AI courses, scatter plots to illustrate clusters within the course dataset, and comparative analysis across continents.

The Scatter Plot Clustering method involves several steps to analyze and visualize course clustering across universities. First, the Term Frequency-Inverse Document Frequency (TF-IDF) technique is used to convert the textual data of course names into numerical features. The TF-IDF score for a term \( t \) in a document \( d \) is given by 
\begin{equation}
    \text{TF-IDF}(t, d) = \text{TF}(t, d) \times \log \left( \frac{N}{\text{DF}(t)} \right)
\end{equation}
where \(\text{TF}(t, d)\) is the term frequency of \( t \) in document \( d \), \(\text{DF}(t)\) is the document frequency of term \( t \) (number of documents containing \( t \)), and \( N \) is the total number of documents. To reduce the dimensionality of the TF-IDF matrix and capture the most significant features, Truncated Singular Value Decomposition (SVD) is applied, which reduces the feature space to 50 dimensions, making subsequent clustering more efficient. Mathematically, SVD decomposes the matrix \( \mathbf{X} \) as \(\mathbf{X} \approx \mathbf{U} \mathbf{\Sigma} \mathbf{V}^T\), where \( \mathbf{U} \) and \( \mathbf{V} \) are orthogonal matrices and \( \mathbf{\Sigma} \) is a diagonal matrix with singular values. The reduced feature matrix is then clustered using the KMeans algorithm with the number of clusters \( k \) set to 8. The algorithm partitions the data into \( k \) clusters by minimizing the within-cluster sum of squares (WCSS): 
\begin{equation}
    \text{WCSS} = \sum_{i=1}^{k} \sum_{x \in C_i} \| x - \mu_i \|^2
\end{equation}
where \( C_i \) is the \( i \)-th cluster, \( \mu_i \) is the centroid of the \( i \)-th cluster, and \( x \) is a data point in cluster \( C_i \). For interpretability, the top 10 words with the highest TF-IDF scores are extracted for each cluster, based on the order of the cluster centroids' feature importance. The t-Distributed Stochastic Neighbor Embedding (t-SNE) technique is then used to visualize the high-dimensional data in a 2D space. This technique reduces the dimensionality while preserving the local structure of the data. The optimization process minimizes the Kullback-Leibler divergence between the joint probabilities of the high-dimensional data and the low-dimensional embedding: 
\begin{equation}
    KL(P \| Q) = \sum_{i \neq j} p_{ij} \log \left( \frac{p_{ij}}{q_{ij}} \right)
\end{equation}
where \( p_{ij} \) and \( q_{ij} \) are the probabilities of data points \( i \) and \( j \) being neighbors in the high-dimensional and low-dimensional spaces, respectively. The clusters are visualized in a scatter plot using the 2D t-SNE coordinates. Each point represents a course, color-coded by its cluster label. 

To complement the detailed methodology for Scatter Plot Clustering, a similar approach is employed to visualize the distribution of AI courses across continents using two different tokenizers.
For the TF-IDF tokenizer, course names are converted into TF-IDF vectors, capturing the importance of each term relative to the entire corpus. These vectors undergo dimensionality reduction via Truncated SVD to retain essential information while minimizing the number of features. KMeans clustering is applied to identify course clusters, and TSNE is used for further dimensionality reduction to visualize the data in a 2D space. 
In contrast, the BERT tokenizer employs a deep contextualized word embedding model to tokenize course names into dense vectors that represent the semantic meaning of each term within the context of the sentence. These vectors are projected onto a lower-dimensional space using TSNE for visualization. KMeans clustering is then applied to identify clusters of courses, and the scatter plot is generated with courses color-coded by continent.
\section{Results}
This section presents the analytical results derived from the collected data, offering insights into the educational landscape of AI programs across the selected universities. The analysis includes a detailed examination of curriculum structures, identification of core AI topics, and a comparative analysis of AI education across different regions.
\subsection{Analysis of Curriculum Structures} 
Most AI-related undergraduate programs are structured as bachelor's degree programs, designed to equip students with foundational knowledge and practical skills in AI. Additionally, some programs offer specialized tracks or concentrations within the broader field of AI, serving as major pathways for students pursuing careers in AI research or related areas such as Computer Science (CS).

To comprehend the academic structure of each university's curriculum, we utilize a stacked bar chart, depicting the distribution of courses across diverse categories for each institution. As depicted in Figure \ref{fig: category}, several notable observations emerge from the analysis:
\begin{figure*}[ht] 
    \centering
    \includegraphics[width=0.9\textwidth]{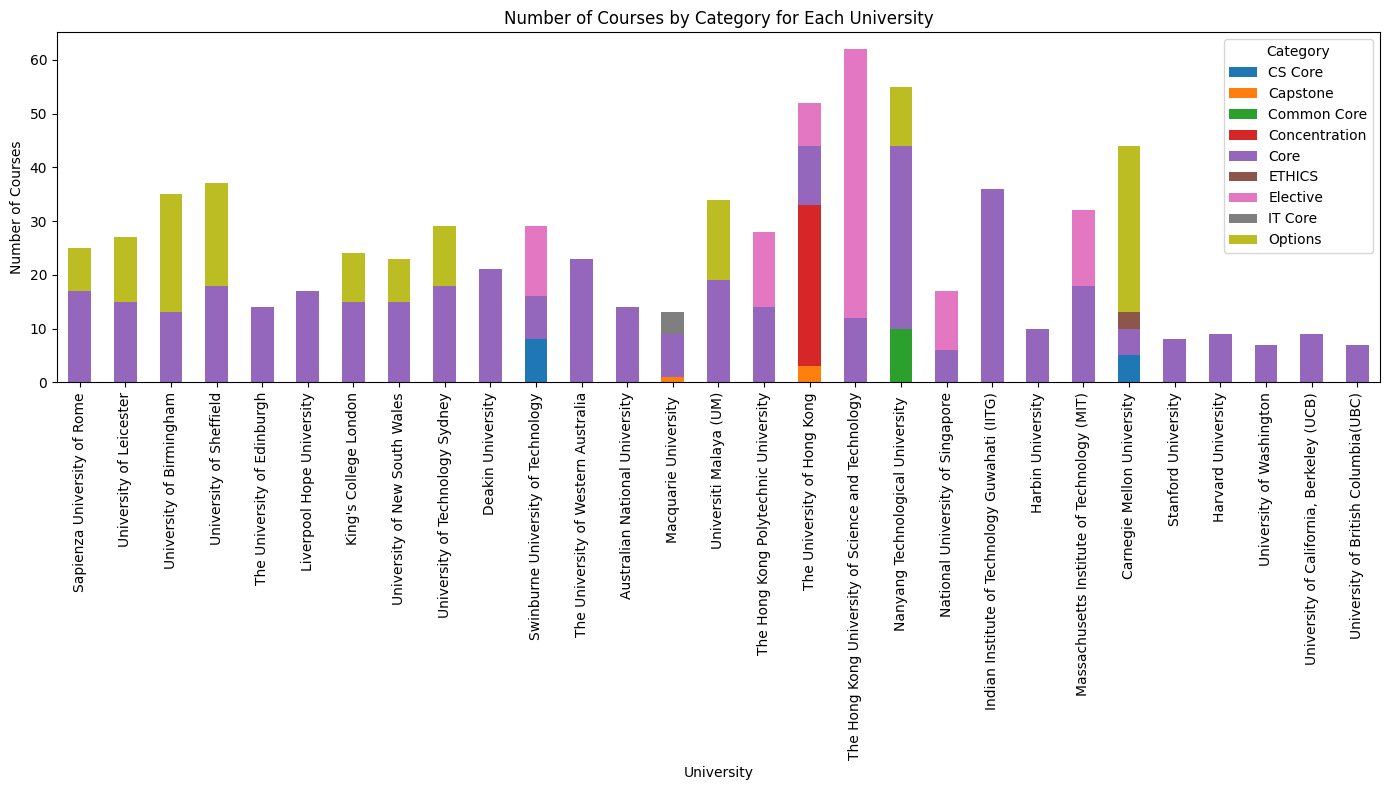} 
    \caption{Stacked Bar Chart Showing the Distribution of Courses by Category Across Different Universities}
    \label{fig: category}
\end{figure*}

Across all institutions, core units in AI education programs are depicted as compulsory requirements, core modules, or essential units for understanding foundational knowledge in this field.
These core units are represented in the stacked bar chart as the purple portion of the bars, indicating their universal inclusion in the curriculum.
In addition, most programs offer optional units in AI, referred to as ``options" or ``electives," providing students with flexibility and choice in their course selection.
For example, the University of Hong Kong offers different concentrations covering multiple units to explore interconnected subjects such as AI in Business Technology and Finance, and AI in Smart City.
This variety of optional units allows students to tailor their education to their interests and career aspirations, fostering specialization in specific areas of AI.

While some universities offer comprehensive AI programs with a wide range of core and optional units, others may only refer to a few core units.
This variation in program structure may be attributed to differences in institutional focus, program objectives, or the integration of AI into broader academic disciplines.
In the case of universities like the University of California, Berkeley (UCB) and the University of Washington, their AI programs primarily focus on core units related to AI research, reflecting a specialized approach to the subject matter. Similarly, institutions such as Stanford University and Harvard University offer specialized programs like the Artificial Intelligence Professional Certificate Program, which may emphasize core units tailored to specific professional objectives rather than offering a broad range of elective options.
Furthermore, universities like the University of British Columbia (UBC) may structure their AI education as certificate or micro-certificate programs, limiting the curriculum to core units necessary for certification. This streamlined approach aligns with the program's objective of providing targeted training in AI skills within a shorter timeframe.

Overall, the analysis underscores the universal presence of core units in AI education programs, alongside the provision of optional units for specialization. The varying approaches to AI education reflect the diverse needs and objectives of students and institutions, resulting in different program structures tailored to meet specific academic, professional, and certification requirements.

\begin{figure*}[ht] % Place the figure at the top of the page
    \centering
    \includegraphics[width=1.0\textwidth]
    {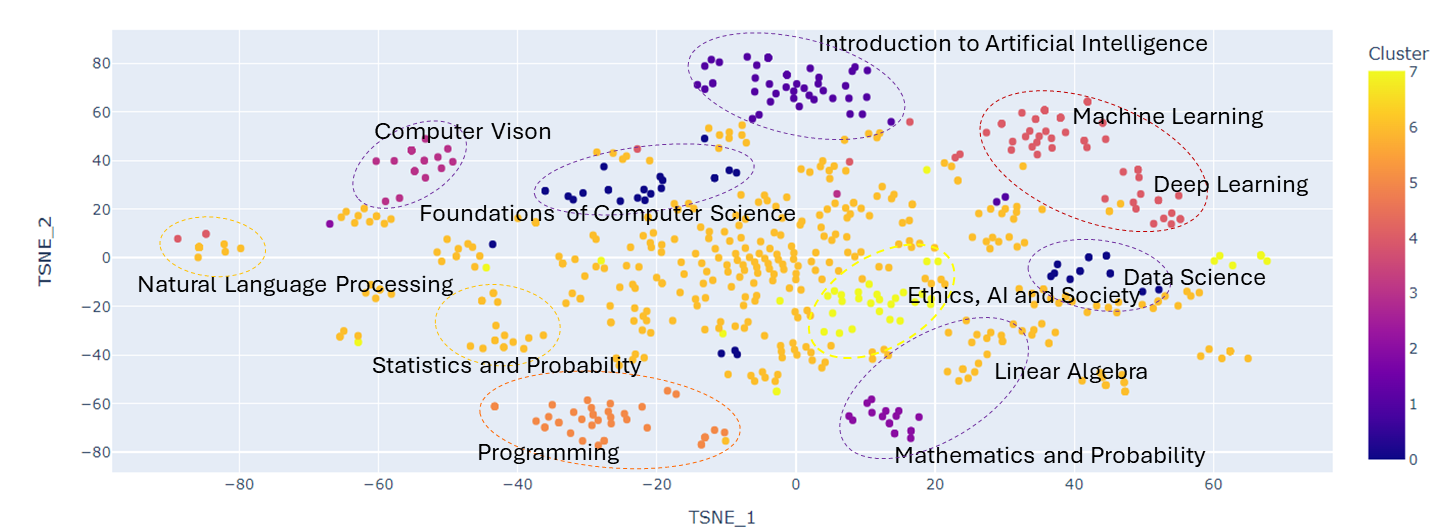}
    \caption{Scatter Plot for Clustered Focus Courses from Different Universities}
    \label{fig: scatter}
\end{figure*}

\begin{figure}[h!]
  \centering
\includegraphics[width=0.45\textwidth]{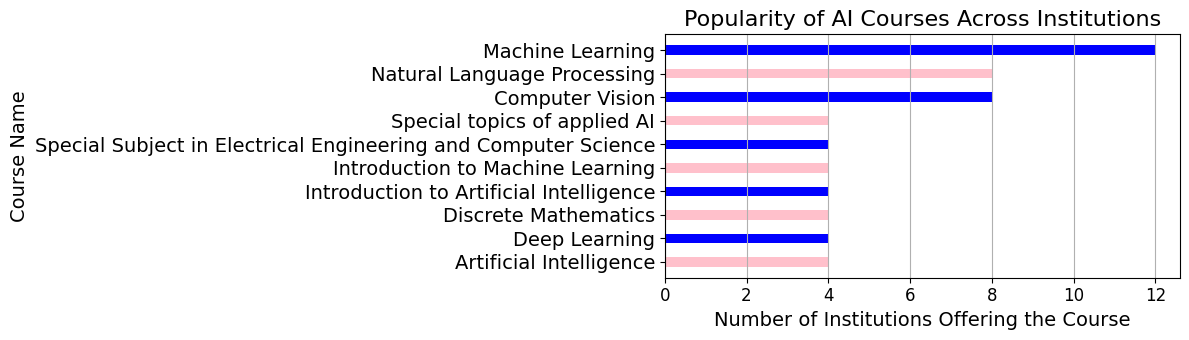}
  \caption{Popularity of AI Courses Across Institutions}
  \label{fig: popularity}
\end{figure}

Apart from the bar chart, we focus on the scatter plot, which offers further insights. 
This plot is generated using all course names from different categories and reveals common focuses among the core and elective units in AI undergraduate education. (Figure \ref{fig: scatter}).
Based on this scatter plot, which visualizes the clustering of courses among different universities, several key observations can be made:
\begin{itemize}
    \item Focus on Fundamental AI Concepts:
    The largest cluster located on the top right of the figure suggests a strong emphasis on fundamental AI concepts across universities. Keywords such as ``Introduction to AI," ``Fundamentals of AI," and ``AI Theory" dominate this cluster, indicating a focus on foundational issues in AI education.
    \item Programming-Related Courses:
    The scatter plot reveals a significant cluster situated in the bottom left quadrant, indicating a concentration of programming-related courses. Notable examples within this cluster encompass ``Python Programming," ``Game Programming," and ``Structured Programming," underscoring the pivotal role of programming as a requisite skill for AI students.
    \item Machine Learning and Deep Learning Foundations:
    In the upper top section of the plot, two clusters coalesce to represent machine learning and deep learning courses, forming a cohesive entity. Prevalent keywords such as ``Machine Learning," ``Deep Learning," and ``Applied Deep Learning" underscore the widespread inclusion of these subjects across university curricula, emphasizing their foundational significance in AI education.
    \item Mathematical Requirements:
    The clusters in the lower right of the graph (labeled 2 and 6), are associated with mathematical concepts such as ``Mathematics," ``Probability," and ``Statistics." This suggests a strong emphasis on the mathematical prerequisites necessary for studying AI.
    \item Specialized AI Subfields: Additional clusters representing specialized AI subfields are scattered throughout the plot. Clusters for ``Natural Language Processing (NLP)," ``Ethics/AI and Society," ``Computer Vision," and "Robotics" indicate a diverse range of course offerings covering various applications and societal implications of AI, underscoring an increased emphasis on the practical applications of AI and their societal and ethical significance.
\end{itemize}
In addition to the course name clusters, the popularity of different AI courses among institutions can be visualized using a bar chart (Figure \ref{fig: popularity}). This chart would display each AI course on the y-axis and the number of institutions offering each course on the x-axis.
Machine learning emerges as the most commonly offered course, with a total of 16 out of 29 institutions providing it, including both the first and sixth most popular courses. Following closely are natural language processing and computer vision, each offered by 8 universities. This trend underscores the significance of these domains in AI education. Additionally, programming-related courses such as ``Introduction to Programming" and fundamental issues like ``Fundamentals of AI" are prevalent, aligning well with the guidance outlined in CS2023. These findings reflect the emphasis on foundational concepts and practical skills in AI curricula across institutions.

\subsection{Identification of Core AI Topics}

To gain insights into the core components of undergraduate AI education across different universities and compare
them with the AI knowledge areas outlined in
the CS2023 curriculum guidance, we have collected the names of core units from various undergraduate programs. These core units represent the foundational pillars upon which AI education is built, encompassing essential topics and concepts integral to understanding and applying AI principles.

In this study, we present a word cloud generated from the compilation of core unit names from diverse undergraduate AI programs (Figure \ref{fig: core cloud}). This word cloud provides a comprehensive overview of the core units taught in AI undergraduate education, offering valuable insights into the common themes and focal areas across different universities. By visually representing the frequency and prominence of core unit names, the word cloud sheds light on the essential components of AI education and highlights the diverse approaches taken by institutions in structuring their undergraduate AI curricula. Observations from the generated word cloud are as follows:
\begin{figure*}[ht] % Place the figure at the top of the page
    \centering
    \includegraphics[width=0.8\textwidth]{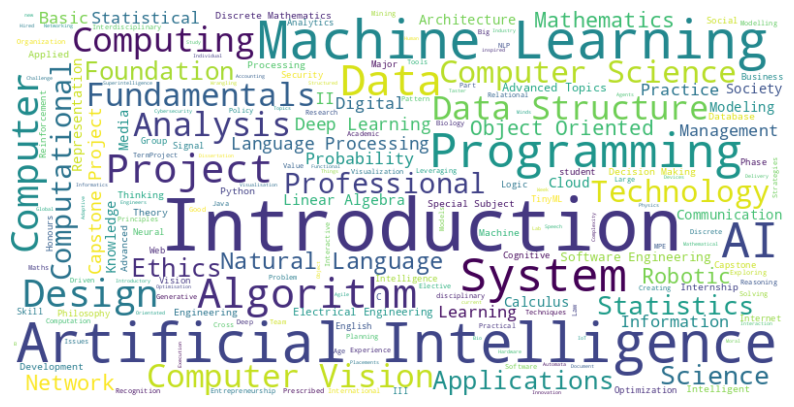} 
    \caption{Word Cloud of Core Units}
    \label{fig: core cloud}
\end{figure*}

\begin{itemize}
    \item \textbf{Prominence of Foundational Concepts:} The word cloud prominently features terms like ``machine learning", ``Programming", ``Algorithm", ``Introduction" and ``fundamentals," indicating their central role in AI education. These terms likely represent core courses that lay the groundwork for understanding AI principles, algorithms, and methodologies as recommended in CS2023.
    \item \textbf{Coverage of Application-Focused Units:} While application-focused units may not dominate the cloud, their presence is notable, indicating a comprehensive coverage of AI applications across the surveyed universities. Their representation across multiple universities indicates a recognition of the importance of practical skills and knowledge in applying AI techniques to real-world problems. These units likely cover topics such as AI in healthcare, finance, robotics, or other specific domains, aligning with CS2023's emphasis on the application of AI techniques in diverse contexts.
    \item \textbf{Presence of Elective Units:} Additionally, units such as ``computer vision" and ``natural language processing" stand out, although to a lesser extent compared to the basic concept. Their appearance suggests that while not required as core units in the CS2023 Artificial Intelligence Knowledge Area, these topics are recognized as important units within AI curricula. Their inclusion reflects a focus on specialized areas of AI, allowing students to deepen their knowledge and expertise based on their interests and career aspirations.
\end{itemize}
Overall, the word cloud provides a snapshot of AI education that aligns with the CS2023 guidance, encompassing a balance of foundational knowledge and practical applications. The distribution of terms suggests a holistic approach to AI education, aiming to equip students with a broad understanding of core principles while also offering opportunities for in-depth exploration of specialized topics.
This alignment with CS2023 ensures that graduates possess the requisite knowledge and skills to contribute effectively to the evolving field of artificial intelligence in various professional settings.

\subsection{Comparative Analysis of AI Education}

In this section, we aim to conduct a comparative analysis across different continents to gain insights into the similarities and differences in the distribution of AI courses.
Firstly, we employ the aforementioned stacked bar chart to stack various categories of courses from each university, facilitating the visualization and comparison of curriculum patterns across different continents.

\begin{figure}[ht!]
  \centering
\includegraphics[width=0.45\textwidth]{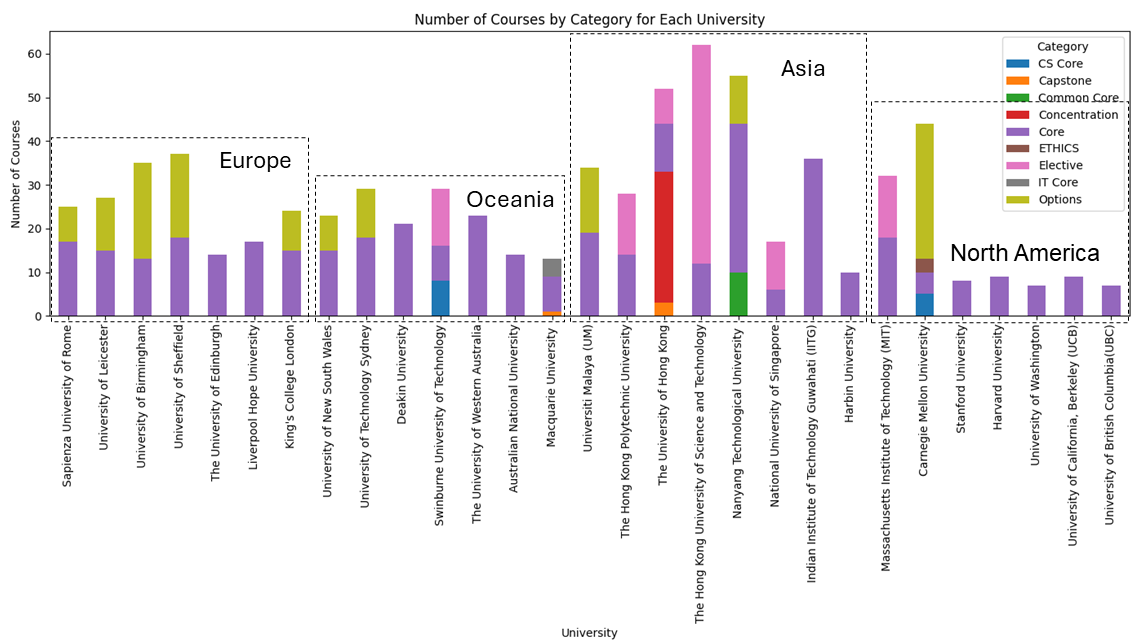}
  \caption{Stacked Bar Chart Illustrating the Distribution of Courses by Category Across Universities in Different Continents}
  \label{fig: stackbar-continents}
\end{figure}

From Figure \ref{fig: stackbar-continents}, we can find that Asian universities stand out with the highest number of courses and variety in their curriculum structures. Specifically, six out of eight Asian universities provide more than 28 courses, reflecting a rich and diverse curriculum.
Contrasting the Asian universities' rich curriculum offerings, universities from other continents, particularly North America, present a contrasting pattern with a lower average number of courses. For example, Oceania universities offer an average of around 21 courses, while those in the North America provide approximately 16 courses on average. This disparity can be attributed to the fact that the majority (6 out of 8) of programs offered by North American universities focus on AI research-related courses or certificate programs.

Universities in Europe, including the UK and Italy, showcase a distinctive trend in offering AI bachelor's degrees with a consistent four-year structure. Most universities in this region follow a uniform model, featuring a blend of compulsory core units alongside elective options for students. 
However, there are notable exceptions to this trend. Liverpool Hope University stands out by offering only mandatory units, without any AI related elective choices. Conversely, The University of Edinburgh adopts a unique approach by offering a broader range of electives across other schools, in addition to mandatory units, enhancing the interdisciplinary aspect of the program.
This comparison highlights the significant differences in course offerings among universities across continents, with Asian universities leading in terms of course variety and richness in their curriculum structures.
This observation aligns well with the result of ``2024 AI Index Report" \cite{stanford2024}. The report shows that the UK has more AI study programs despite having fewer universities overall compared to the US based on the \cite{studyportals}. (In 2023, the UK had 744 English-language AI study programs, while the US had 667.) The factors like Studyportals' coverage and the structure of higher education could definitely contribute to this. In the UK, there might be a more concentrated effort to offer specialized AI programs, whereas in the US, students interested in AI might pursue it within broader computer science programs. This nuanced understanding helps paint a clearer picture of AI education across different countries.

\begin{figure*}[ht!]
  \centering
\includegraphics[width=0.9\textwidth]{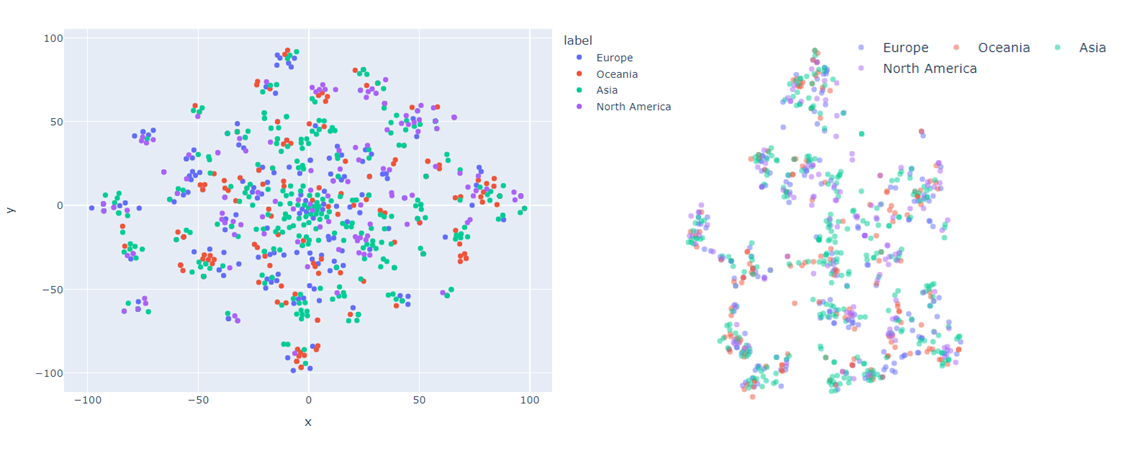}
  \caption{Scatter Plots of Courses across Continents (Tokenizers: Left/TF-IDF, Right/BERT)}
  \label{fig:scatter_comparison}
\end{figure*}

Furthermore, we can visually compare the distinct patterns of course distribution across continents in the context of AI undergraduate education through Scatter Plots of Courses, as depicted in Figure \ref{fig:scatter_comparison}. This figure showcases scatter plots generated using two different tokenization techniques.
The resulting scatter plot color-codes courses based on the continent they belong to, enabling a comparative analysis of course distributions across geographical regions.
The scatter plot generated using the TF-IDF tokenizer reveals a dispersed pattern of course distributions across Asia and Europe. While these continents exhibit some dense clusters, indicating a diverse range of course offerings, there is also a broader dispersion of courses, suggesting a broader coverage of AI topics.
In contrast, universities in North America and Oceania demonstrate a more focused distribution, with fewer dense clusters. This concentration of courses suggests a focus on specific areas within AI education rather than a broader coverage of topics.
Similarly, the scatter plot generated using the BERT tokenizer also showcases a dispersed pattern of course distributions across Asia and Europe. However, the dispersion tends to cover a broader area, indicating a more diverse range of offerings.
Universities in North America and Oceania exhibit a focused distribution, with dense clusters in specific areas. This focus suggests a prioritization of particular topics or domains within AI education.

Comparing the scatter plots generated using TF-IDF and BERT tokenizers, we observe consistent patterns across continents. Both tokenizers reveal dispersed distributions in Asia and Europe, indicating a diverse range of course offerings, while North America and Oceania exhibit more focused distributions. This suggests that universities in these regions prioritize specific areas within AI education, with varying degrees of breadth and depth in their course offerings.
The analysis of scatter plots generated using different tokenizers provides valuable insights into the distribution of AI courses across continents. By comparing patterns observed in TF-IDF and BERT tokenizers, we gain a nuanced understanding of regional emphases and priorities in AI undergraduate education, highlighting areas of diversity and specialization across geographical regions.
\section{Discussion}
In addition to the findings presented in the Results section, we also have important observations that warrant further discussion. This section will delve into the analysis of math requirements and explore the career paths available for AI graduates. By examining these aspects, we aim to provide a comprehensive understanding of the educational frameworks and their alignment with industry demands and career opportunities in AI.
\subsection{Analysis of Math Requirements}
Based on our observations, most AI undergraduate programs around the world require students to have advanced-level qualifications in mathematics as part of their entry requirements. For example, the University of Birmingham in the UK requires students to include A-level Mathematics with a grade A, while the University of Technology Sydney in Australia requires Higher School Certificate (HSC) Mathematics Advanced.

In addition to the mathematical entry requirements, students are typically required to take compulsory units related to mathematics within AI undergraduate programs. Among 29 institutions analyzed, it was found that:
21 courses include ``mathematics", indicating a strong emphasis on mathematical concepts and principles.
13 courses include ``statistics", highlighting the importance of statistical analysis and inference in AI applications.
12 courses include the word ``probability", indicating the significance of probability theory in understanding uncertainty and making informed decisions in AI.
12 courses include ``calculus", underscoring the necessity of calculus in modeling and optimizing algorithms and systems in AI.
Overall, these findings suggest that AI undergraduate programs place a significant emphasis on mathematical proficiency, with students expected to have a strong foundation in mathematical principles upon entry and to further develop their mathematical skills through compulsory math-related courses throughout the program. This emphasis reflects the central role that mathematics plays in the theory, development, and application of artificial intelligence.
\subsection{Career Paths for AI Graduates}
As the field of Artificial Intelligence (AI) continues to expand and evolve, so do the career opportunities for graduates of AI undergraduate programs. With a solid foundation in machine learning, computer vision, data science, and other AI-related disciplines, graduates are well-equipped to pursue a wide range of career paths.
As most institutions suggest, AI graduates often find rewarding careers as data scientists, given the strong overlap between AI and data science disciplines. With their expertise in machine learning algorithms, statistical analysis, and programming, AI graduates are well-suited to extract valuable insights from large datasets and drive data-driven decision-making across various industries.

Following closely behind data scientist roles, AI graduates frequently pursue careers as machine learning engineers, leveraging their skills to develop and deploy machine learning models that power intelligent systems and applications. Additionally, computer vision engineers are in high demand, as industries increasingly rely on image recognition technology for tasks such as facial recognition, object detection, and augmented reality applications.

Furthermore, AI graduates often succeed in roles such as software engineers, developers, and IT consultants, where their AI expertise enhances product development, system optimization, and technology implementation. Additionally, opportunities abound in cloud computing engineering and cybersecurity analysis, as organizations seek to leverage AI technologies to enhance security and scalability in cloud-based environments.
As the AI field continues to grow, specialized roles such as AI ethicists, AI architects, and AI policy advisors are emerging to address the ethical, legal, and societal implications of AI technologies. Graduates with a strong foundation in AI principles and ethics are well-positioned to shape the future of AI responsibly and sustainably.

Overall, AI graduates have a wealth of career opportunities, ranging from traditional roles in data science and machine learning to emerging roles in AI ethics, policy, and governance. By leveraging their skills and expertise, AI graduates can make significant contributions to the advancement and responsible deployment of AI technologies across various domains.

\section{Conclusion}
The vision of AI education globally reflects a balance between foundational knowledge and specialized skills, tailored to diverse academic and professional objectives. Our research collected 750 course names covered by 29 universities' undergraduate AI education programs, providing a comprehensive dataset for analysis. Firstly, the analysis of curriculum structures highlights the universal presence of core units, with optional units for specialization. Machine learning, computer vision, and natural language processing emerge as key courses, emphasizing their significance in AI education. Secondly, the identification of core AI topics aligns with CS2023 guidelines, ensuring graduates are well-equipped with essential knowledge and practical skills for professional success. Lastly, a comparative analysis across continents reveals that Asian and European universities offer a diverse range of courses, while North American and Oceanian institutions focus on specific subfields. These patterns underscore regional differences and priorities, offering a thorough comparative analysis and a comprehensive understanding of global AI education trends. 
Our research provides significant insights into the diverse approaches and regional specializations that shape the structure and focus of AI education worldwide."

Future research should explore the longitudinal evolution of AI curricula to understand how educational programs adapt to emerging technologies and industry demands. Additionally, examining the impact of these educational structures on student outcomes, such as employability and innovation capacity, will provide deeper insights into the effectiveness of various curriculum designs. By continuing to analyze and refine AI educational programs, we can better prepare students for the dynamic and rapidly evolving field of artificial intelligence.

\section*{Declaration of Conflicting Interests}
The author(s) declared no potential conflicts of interest with respect to the research, authorship, and/or publication of this article.
\section*{Funding}
The author(s) received no financial support for the research, authorship, and/or publication of this article.

\bibliographystyle{sageh}
\bibliography{Reference.bib}

\begin{thebibliography}{15}
\providecommand{\natexlab}[1]{#1}
\providecommand{\url}[1]{\texttt{#1}}
\providecommand{\urlprefix}{URL }
\expandafter\ifx\csname urlstyle\endcsname\relax
  \providecommand{\doi}[1]{DOI:\discretionary{}{}{}#1}\else
  \providecommand{\doi}{DOI:\discretionary{}{}{}\begingroup \urlstyle{rm}\Url}\fi

\bibitem[{Brusilovsky(2024)}]{Brusilovsky24}
Brusilovsky P (2024) {AI} in education, learner control, and human-ai collaboration.
\newblock \emph{Int. J. Artif. Intell. Educ.} 34(1): 122--135.
\newblock \doi{10.1007/S40593-023-00356-Z}.
\newblock \urlprefix\url{https://doi.org/10.1007/s40593-023-00356-z}.

\bibitem[{Doughty et~al.(2024)Doughty, Wan, Bompelli, Qayum, Wang, Zhang, Zheng, Doyle, Sridhar, Agarwal, Bogart, Keylor, K{\"{u}}lt{\"{u}}r, Savelka and Sakr}]{DoughtyWBQWZZDS24}
Doughty J, Wan Z, Bompelli A, Qayum J, Wang T, Zhang J, Zheng Y, Doyle A, Sridhar P, Agarwal A, Bogart C, Keylor E, K{\"{u}}lt{\"{u}}r C, Savelka J and Sakr M (2024) A comparative study of ai-generated {(GPT-4)} and human-crafted mcqs in programming education.
\newblock In: Herbert N and Seton C (eds.) \emph{Proceedings of the 26th Australasian Computing Education Conference, {ACE} 2024, Sydney, NSW, Australia, 29 January 2024- 2 February 2024}. {ACM}, pp. 114--123.
\newblock \doi{10.1145/3636243.3636256}.
\newblock \urlprefix\url{https://doi.org/10.1145/3636243.3636256}.

\bibitem[{Druga et~al.(2022)Druga, Otero and Ko}]{Druga22}
Druga S, Otero N and Ko AJ (2022) The landscape of teaching resources for ai education.
\newblock In: \emph{Proceedings of the 27th ACM Conference on on Innovation and Technology in Computer Science Education Vol. 1}, ITiCSE '22. New York, NY, USA: Association for Computing Machinery.
\newblock ISBN 9781450392013, p. 96–102.
\newblock \doi{10.1145/3502718.3524782}.
\newblock \urlprefix\url{https://doi-org.ezproxy-b.deakin.edu.au/10.1145/3502718.3524782}.

\bibitem[{Eaton and Epstein(2024)}]{EatonE24}
Eaton E and Epstein SL (2024) Artificial intelligence in the {CS2023} undergraduate computer science curriculum: Rationale and challenges.
\newblock In: Wooldridge MJ, Dy JG and Natarajan S (eds.) \emph{Thirty-Eighth {AAAI} Conference on Artificial Intelligence, {AAAI} 2024, Thirty-Sixth Conference on Innovative Applications of Artificial Intelligence, {IAAI} 2024, Fourteenth Symposium on Educational Advances in Artificial Intelligence, {EAAI} 2014, February 20-27, 2024, Vancouver, Canada}. {AAAI} Press, pp. 23078--23083.
\newblock \doi{10.1609/AAAI.V38I21.30352}.
\newblock \urlprefix\url{https://doi.org/10.1609/aaai.v38i21.30352}.

\bibitem[{Garvey and Patrick(2023)}]{10371554}
Garvey and Patrick G (2023) Artificial intelligence and art: A university curriculum course for undergraduates.
\newblock In: \emph{2023 14th IIAI International Congress on Advanced Applied Informatics (IIAI-AAI)}. pp. 667--670.
\newblock \doi{10.1109/IIAI-AAI59060.2023.00131}.

\bibitem[{Gibellini et~al.(2023)Gibellini, Fabretti and Schiavo}]{Gibellini23}
Gibellini G, Fabretti V and Schiavo G (2023) Ai education from the educator’s perspective: Best practices for an inclusive ai curriculum for middle school.
\newblock In: \emph{Extended Abstracts of the 2023 CHI Conference on Human Factors in Computing Systems}, CHI EA '23. New York, NY, USA: Association for Computing Machinery.
\newblock ISBN 9781450394222.
\newblock \doi{10.1145/3544549.3585747}.
\newblock \urlprefix\url{https://doi-org.ezproxy-b.deakin.edu.au/10.1145/3544549.3585747}.

\bibitem[{Nithithanatchinnapat et~al.(2024)Nithithanatchinnapat, Maurer, Deng and Joshi}]{Nithithanatchinnapat24}
Nithithanatchinnapat B, Maurer J, Deng XN and Joshi KD (2024) Future business workforce: Crafting a generative ai-centric curriculum today for tomorrow's business education.
\newblock \emph{SIGMIS Database} 55(1): 6–11.
\newblock \doi{10.1145/3645057.3645059}.
\newblock \urlprefix\url{https://doi-org.ezproxy-b.deakin.edu.au/10.1145/3645057.3645059}.

\bibitem[{{Quacquarelli Symonds (QS)}(2024)}]{qs2024}
{Quacquarelli Symonds (QS)} (2024) {QS World University Rankings by Subject 2024: Data Science and Artificial Intelligence}.
\newblock \url{https://www.topuniversities.com/university-subject-rankings/data-science-artificial-intelligence?page=0&tab=indicators&sort_by=rank&order_by=asc}.

\bibitem[{{Stanford University}(2024)}]{stanford2024}
{Stanford University} (2024) {2024 AI Index Report}.
\newblock \urlprefix\url{https://aiindex.stanford.edu/report/}.

\bibitem[{{Studyportals}(2024)}]{studyportals}
{Studyportals} (2024) {Studyportals}.
\newblock \url{https://studyportals.com/}.
\newblock Accessed: 2024-05-08.

\bibitem[{Wang and Yu(2023)}]{Wang23}
Wang Y and Yu F (2023) Visualization and analysis of mapping knowledge domains for ai education system studies.
\newblock In: \emph{2023 11th International Conference on Information and Education Technology (ICIET)}. pp. 475--479.
\newblock \doi{10.1109/ICIET56899.2023.10111115}.

\bibitem[{Williamson(2024)}]{williamson2024}
Williamson B (2024) The social life of ai in education.
\newblock \emph{Int J Artif Intell Educ} 34: 97--104.
\newblock \doi{10.1007/s40593-023-00342-5}.
\newblock \urlprefix\url{https://doi.org/10.1007/s40593-023-00342-5}.

\bibitem[{Xu et~al.(2024)Xu, Du, Niyato, Kang, Xiong, Mao, Han, Jamalipour, Kim, Shen, Leung and Poor}]{Xu24}
Xu M, Du H, Niyato D, Kang J, Xiong Z, Mao S, Han Z, Jamalipour A, Kim DI, Shen X, Leung VCM and Poor HV (2024) Unleashing the power of edge-cloud generative ai in mobile networks: A survey of aigc services.
\newblock \emph{IEEE Communications Surveys \& Tutorials} 26(2): 1127--1170.
\newblock \doi{10.1109/COMST.2024.3353265}.

\bibitem[{Zhou et~al.(2021)Zhou, Tong, Lan, Zheng and Zhan}]{Zhou2021}
Zhou X, Tong Y, Lan X, Zheng K and Zhan Z (2021) Ai education in massive open online courses: A content analysis.
\newblock In: \emph{2021 3rd International Conference on Computer Science and Technologies in Education (CSTE)}. pp. 80--85.
\newblock \doi{10.1109/CSTE53634.2021.00023}.

\bibitem[{Zhou et~al.(2022)Zhou, Zhan, Liu, Wan, Liu and Zou}]{ZhouZLWLZ22}
Zhou Y, Zhan Z, Liu L, Wan J, Liu S and Zou X (2022) International prospects and trends of artificial intelligence education: {A} content analysis of top-level {AI} curriculum across countries.
\newblock In: \emph{Proceedings of the 6th International Conference on Digital Technology in Education, {ICDTE} 2022, Hangzhou, China, September 16-18, 2022}. {ACM}, pp. 337--343.
\newblock \doi{10.1145/3568739.3568796}.
\newblock \urlprefix\url{https://doi.org/10.1145/3568739.3568796}.

\end{thebibliography}

\end{document}